# RAMPART: RowHammer Mitigation and Repair for Server Memory Systems

Steven C. Woo, Wendy Elsasser, Mike Hamburg, Eric Linstadt,
Michael R. Miller, Taeksang Song, James Tringali
Rambus Labs, Rambus Inc., San Jose, CA
{swoo,welsasser,mhamburg,elinstadt,michaelm,tsong,jamestr}@rambus.com

## Abstract

RowHammer attacks are a growing security and reliability concern for DRAMs and computer systems as they can induce many bit errors that overwhelm error detection and correction capabilities. System-level solutions are needed as process technology and circuit improvements alone are unlikely to provide complete protection against RowHammer attacks in the future. This paper introduces RAMPART, a novel approach to mitigating RowHammer attacks and improving server memory system reliability by remapping addresses in each DRAM in a way that confines RowHammer bit flips to a single device for any victim row address. When RAMPART is paired with Single Device Data Correction (SDDC) and patrol scrub, error detection and correction methods in use today, the system can detect and correct bit flips from a successful attack, allowing the memory system to heal itself. RAMPART is compatible with DDR5's RowHammer mitigation features, as well as a wide variety of algorithmic and probabilistic tracking methods. We also introduce BRC-VL, a variation of DDR5's Bounded Refresh Configuration (BRC) that improves system performance by reducing mitigation overhead and show that it works well with probabilistic sampling methods to combat traditional and victim-focused mitigation attacks like Half-Double. The combination of RAMPART, SDDC, and scrubbing enables stronger RowHammer resistance by correcting bit flips from one successful attack. Uncorrectable errors are much less likely, requiring two successful attacks before the memory system is scrubbed.

## 1. Introduction

DRAMs are increasingly susceptible to RowHammer attacks [1], in which repeated activations of an *aggressor* row can flip bits in nearby *victim* rows. Interest in RowHammer attacks is growing due to the potential for data corruption and security exploits, especially in servers and data centers [2][3]. Although servers have employed Reliability, Availability, and Serviceability (RAS) methods such as symbol-based Error Correction Codes (ECC) for many generations, RowHammer attacks are dangerous because they can cause many errors across multiple DRAMs on a DIMM (Dual In-line Memory Module) that can overwhelm the detection and correction capabilities of ECC. The hammer count *HC*, the number of aggressor row activates needed to cause bit flips in victim rows, has dropped by more than an order of magnitude over the past decade [3], and recent work has discussed the coupling and leakage mechanisms behind RowHammer bit flips [4][5]. DRAM cell vulnerability will continue to get worse at smaller process geometries as reduced cell-to-cell spacing increases susceptibility to disturb induced charge leakage, and increases the number of victim rows. For these reasons, RowHammer is a fundamental scaling problem for future DRAMs that process technology and circuits alone are unlikely to fully solve.

Many attacks have been demonstrated, including some that can result in access to privileged information and denial-of-service [11]-[17]. DRAM RowHammer vulnerability and mitigation methods in use today have been discussed in previous research [2],[7]-[10]. DDR4 Target Row Refresh (TRR) selects aggressor rows and refreshes victims to restore bit cell charge. TRR algorithms are vendor and device-specific and are not public, but these algorithms have been reverse-engineered [2], enabling targeted attacks to be crafted that bypass them. Newer *victim-focused mitigation attacks* like Half-Double [18] also circumvent DDR4 protections [32], illustrating the danger of attacks that have yet to be discovered. DDR5 Refresh Management (RFM) [6] provides protection against RowHammer attacks and is also vendor and device-specific with details that are not public. Obscuring these details makes it impossible to estimate system-level resistance to RowHammer attacks. Many works propose prevention and mitigation methods [19]-[45], with varying area and performance impacts [7] that change with HC [44].

This paper presents RAMPART (*Row Address Map Permutation And Reassignment Technique*), a new approach to RowHammer mitigation in server memory systems that is compatible with DDR5's existing RowHammer mitigation features. RAMPART combines algorithmic row address remapping in the DRAMs together with common RAS methods and leverages existing circuits and process technology. Unlike other remapping methods [32][41][44], remapping only happens once during manufacturing or initialization, so remapping doesn't affect performance. RAMPART mitigates RowHammer attacks by (i) confining bit flips to a single DRAM for any victim row address, and (ii) enabling RowHammer bit flips to be corrected with SDDC, allowing the system to continue running. The combination of RAMPART and SDDC works no matter how many bits flip as a result of a single successful attack, allowing the system to function properly *even if the attack flips every bit in the* *neighboring rows.* When RAMPART is combined with SDDC and patrol scrub, features in use today, uncorrectable errors require a second successful attack before errors from the first attack are detected and corrected, improving the resistance of the memory system.

An important objective of RowHammer mitigation and prevention is to increase the memory system's resistance to data corruption while minimizing any performance impact. RAMPART can be used together with many previously described prevention and mitigation methods, including other remapping techniques, and can be coupled with a variety of algorithmic tracking and probabilistic selection methods. When probabilistic sampling is used, resistance to data corruption can be substantially increased with the combination of RAMPART and SDDC since two successful attacks are needed to irreparably damage data. We also introduce BRC-VL, a variation of DDR5 BRC with lower performance overhead that works well with RAMPART and SDDC, and discuss a practical silicon implementation.

We analyze DDR5-5600 memory systems across a range of HC values that reflect near-term possibilities based on recent studies. For the range of scenarios simulated, our analysis shows that the combination of RAMPART, SDDC, BRC-VL, and patrol scrub can reduce the probability of data corruption by 4 to 17 orders of magnitude after one year of continuous attacks compared to using DDR5 BRC. For the benchmarks and parameters studied, BRC-VL's lower overhead results in a smaller CPI impact versus DDR5 BRC when both are compared to a system running with no RowHammer mitigation. The relative CPI difference can be as high as 3.9% at equal RFM issue rates, and can be higher when targeting equal memory resistance to data corruption.

Our paper makes the following contributions:
- We introduce RAMPART, a novel DRAM row address remapping method that confines RowHammer bit flips to a single DRAM for any victim row address. RAMPART is compatible with DDR5's current mitigation mechanisms, as well as many other mitigation and prevention techniques.
- We show that when RAMPART is coupled with existing ECC methods, bit flips from a successful RowHammer attack can be corrected, and the system can keep running. Uncorrectable errors require two successful attacks before damage from the first attack is detected and corrected.
- We introduce BRC-VL, a variation of DDR5 BRC that improves system performance and resistance to data corruption when coupled with RAMPART and SDDC. Our DDR5-5600 models show the probability of corrupted data can be reduced by 4 to 17 orders of magnitude after one year of continuous attacks for the parameters studied. We also describe a practical TSMC 7nm implementation.

## 2. Background

### 2.1 DRAMs and DIMMs

Figure 1a illustrates the structure of DRAMs and DIMMs. DRAMs store bits of data in arrays of capacitors grouped together in a hierarchical structure that balances storage density and performance. DRAM bit cells are organized into groups called *rows* or *pages*, with multiple rows aggregated into mats, and mats aggregated into banks. Multiple banks form a *bank group* that share some resources. Banks have a 2D organization, with bit cells identified by *row* and *column addresses* that are decoded to locate bits in this structure. Additional rows called *spare* or *redundant rows* are enabled during test and repair to take the place of regular rows that have manufacturing defects. Row addresses for the repaired rows are remapped internally to these spare rows.

A *command decoder* receives commands and addresses from the *memory controller* (or *controller* for short) via a *command/address* (CA) bus connected to the CA pins. When data is read from a DRAM, an *activate* command provides a row address to specify that all bits in one row of a bank are to be moved to *sense amplifiers* (or *sense amps*). A separate *read command* provides a column address to select a subset of the bits from the sense amps to be returned across the DQ pins. Write commands function in a similar manner, providing data to be written into the DRAM.

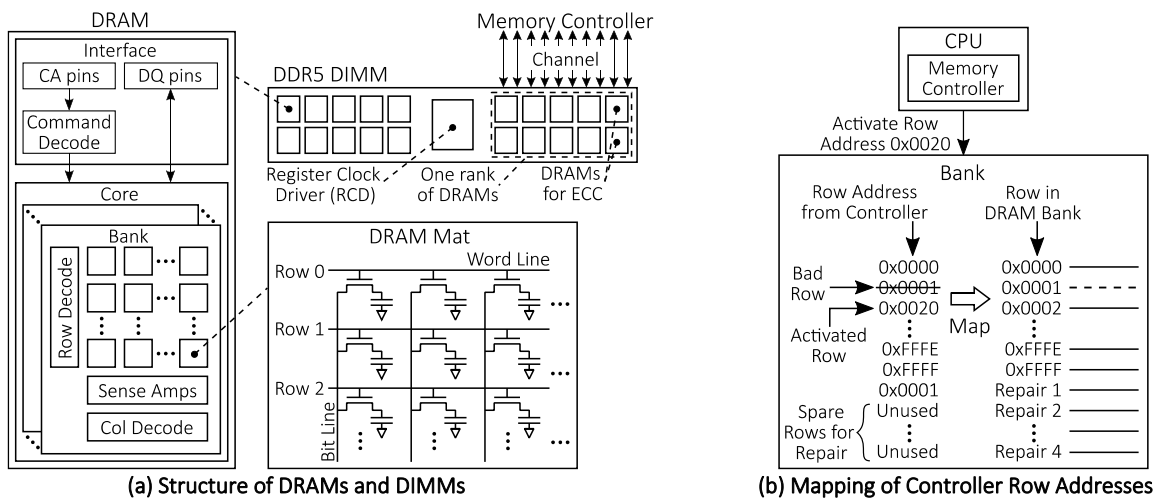

(a) Structure of DRAMs and DIMMs  (b) Mapping of Controller Row Addresses

Figure 1: The structure of DRAMs and DIMMs is shown in (a). A memory channel communicates data between the controller and a rank of DRAMs on a DIMM. Row addresses issued by the controller are mapped to rows in the DRAM banks, as shown in (b). Defective rows are mapped to spare rows during manufacturing. The mapping shown in this example is fictitious.

Data is passed on a *memory channel* between the controller and multiple DRAMs on a DIMM. DDR5 DIMMs contain more than one channel, and the DRAMs associated with a channel are partitioned into independent subsets called *ranks*. One rank of DRAMs responds to a transaction initiated by the controller, with the data spread among the DRAMs in the rank. ECC can be provided with up to 2 additional DRAMs in each rank, allowing a limited number of errors to be corrected. With 2 additional DRAMs, *Single Device Data Correction* (SDDC) is enabled, allowing the memory system to continue operating even if one DRAM in the rank fails or returns data that is completely incorrect. The same row and column addresses are passed to all DRAMs in the rank, and each decodes these addresses in the same way to access data. Processors can connect to multiple memory channels, enabling higher capacities and bandwidths.

Figure 1b illustrates how controller row addresses map to rows in the DRAM banks. The mapping shown is fictitious but is illustrative of the function. Mappings are vendor and product-specific, and are not public [3], making it unclear which rows are neighbors [17]. Each DRAM on a DIMM is from the same manufacturer and maps row addresses in the same way. Address mappings can be reverse engineered [3][7], revealing the mapping algorithm used for a given DRAM product. An important exception is for repaired rows that are remapped to spare rows. Repaired rows can be different in each DRAM, and there is flexibility on where they can be remapped, making the location and neighbors of these rows DRAM-specific. These mappings are not public, making the DRAM the only device that knows the complete mapping of addresses to rows in the DRAM core.

## 2.2 RowHammer in DRAMs and DIMMs

A *traditional* RowHammer attack, shown in Figure 2a, repeatedly activates an *aggressor* row (row R) to cause bits in neighboring *victim* rows (rows R±1) to flip before their data can be refreshed. Double-sided RowHammer attacks repeatedly activate 2 rows (rows R-2 and R) on either side of a victim row (row R-1), making it harder to identify them. At smaller process geometries the hammer count continues to fall, and aggressor rows have a larger *blast radius* that affects victims multiple rows away, enabling more sophisticated attacks. Refreshing victim rows can stop these attacks.

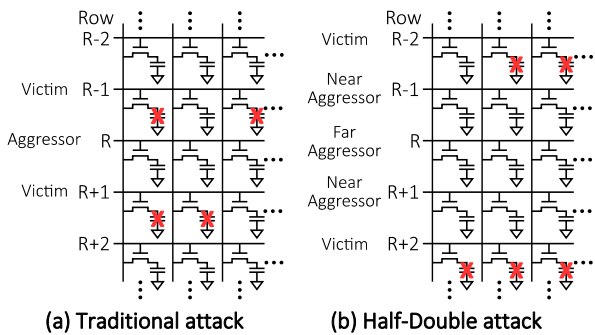

(a) Traditional attack    (b) Half-Double attack

**Figure 2:** Examples of (a) a traditional RowHammer attack and (b) a Half-Double attack. Half-Double attacks use refreshes of Rows R-1 and R+1 (victims of aggressor row R) to act as aggressors on victim Rows R-2 and R+2.

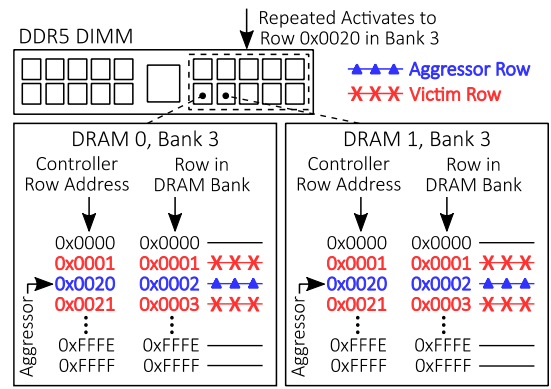

**Figure 3:** DRAMs on a DIMM map controller row addresses to rows in DRAM banks in the same way. A RowHammer attack can flip bits in the same victim row in multiple DRAMs, overwhelming ECC protection.

Half-Double attacks, shown in Figure 2b, flip bits in victim rows that are multiple rows away from the aggressor row. This attack hammers row R, the *far-aggressor*, and relies on refresh operations being issued to rows R±1 that are designed to stop a traditional attack. These refresh operations act as *near-aggressors* that hammer rows that are two rows away (rows R±2) from the far-aggressor. Half-Double attacks use *victim-focused mitigation operations* (refreshes to rows R±1) as hammers, leveraging methods designed to stop traditional attacks [18][32].

Figure 3 illustrates how a RowHammer attack can cause uncorrectable errors in server memory systems. As discussed in Section 2.1, each DRAM maps controller row addresses to internal rows in the same way. An important observation is that when a controller row address is hammered, *the same controller row addresses are victims in each DRAM in the rank*. If multiple DRAMs have errors at the same victim row address, ECC detection mechanisms will be overwhelmed and will fail. In the worst case, many errors in multiple DRAMs will be undetected, leading to silent data corruption.

## 3. RAMPART: Confining Victim Addresses to a Single DRAM

Our solution approach is to confine victim row bit flips for any address to a single DRAM, and to use existing ECC methods to detect and correct these errors. The foundation of our solution is the observation that *bit flips for any victim row address can be confined to a single DRAM if controller row addresses are never neighbors in more than one DRAM*. This can be achieved if each DRAM in the rank implements a unique mapping of controller row addresses to rows within the banks. In this case, a successful RowHammer attack can still flip bits in each DRAM, but they will be in victim rows that correspond to different controller row addresses in each DRAM. By mapping addresses in this way, potential bit flips are spread among different victim row addresses in different DRAMs, in a way that allows ECC to correct them. Reading data from any of these victim row addresses will result in erroneous data from only one DRAM, which can be corrected with existing system-level ECC methods described in Section 4, allowing the system to continue running. Irreparable damage to data requires a second successful attack to the same victim row address, but in a different DRAM, before errors from the first attack are corrected with ECC.

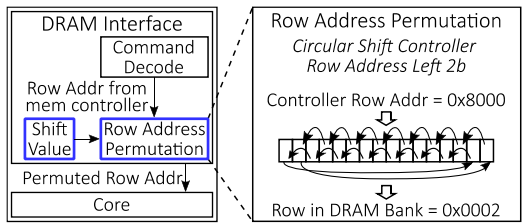

Figure 4: Example RAMPART row address remapping. Controller row addresses are circularly shifted left by 2 bits. The resulting address is the row in the DRAM bank.

In this section we introduce RAMPART, a remapping method enabling different row address mappings in each DRAM so that row addresses are never neighbors in more than one DRAM in the rank. By remapping in this way, RAMPART provides the benefits of (i) confining bit flips to a single DRAM for any victim row address, allowing them to be corrected with SDDC ECC, and (ii) requiring two successful attacks (before bit flips from the first attack are corrected) to cause irreparable damage to data. In addition, remapping is only done once, at manufacturing time or during initialization, and does not affect performance.

Figure 4 illustrates RAMPART using a circular left shift remapping function that takes *controller row addresses* issued by the controller and permutes them so that they map to different rows in the DRAM bank. This example shows that controller row address 0x8000 is circularly shifted left by a programmed *shift value* of 2 bits, and the resulting value 0x0002 is the row in the bank that this controller row address maps to. Other functions are possible, for example using Linear Feedback Shift Registers (LFSRs).

Generalized circular shift hardware can be implemented easily in the DRAM interface with no additional latency penalty, as repaired rows already go through a remapping process. Based on our analysis of die photos such as that shown in [48], DRAM peripheral circuits including the DLL, IO, command decoder, data path and command path occupy about 3% of DRAM area, and an additional 16b circular shifter has no die size impact. Each DRAM in the rank can be programmed with a different shift value based on a unique ID programmed during initialization or set with fuses at manufacturing, for example. Implementing RAMPART remapping in the DRAM also allows the mapping of controller row addresses to core rows to remain hidden from the memory controller.

Figure 5a illustrates a 16b row address mapping (similar in size to DDR5) that assigns each DRAM in the rank a unique ID value, and circularly left shifts controller row addresses by this ID value (1-bit×ID). The tables at the bottom of this figure illustrate the unique mappings that RAMPART provides for each DRAM. In each table, the controller row address is shown in the left column, and the right column shows the row in the bank that this controller row address maps to. In this example, controller row address 0x0000 always maps to the first row in each DRAM, as circularly left shifting 0x0000 by any number of bits always results in 0x0000. But controller row address 0x0001 maps to different rows depending on how many bits it is shifted by. In DRAM 0, controller row address 0x0001 occupies row 0x0001 (shift by 0 bits). In DRAM 1, controller row address 0x0001 is in row 0x0002 (shift by 1 bit). And in DRAM 2, controller row address 0x0001 is in row 0x0004 (shift by 2 bits). The figure shows controller row address 0x0001 is next to controller row address 0x0000 only in DRAM 0 and moves further away in other DRAMs due to the increasing size of the shift. Using this mapping, neighbors of any given row address in one DRAM are never neighbors in another DRAM.

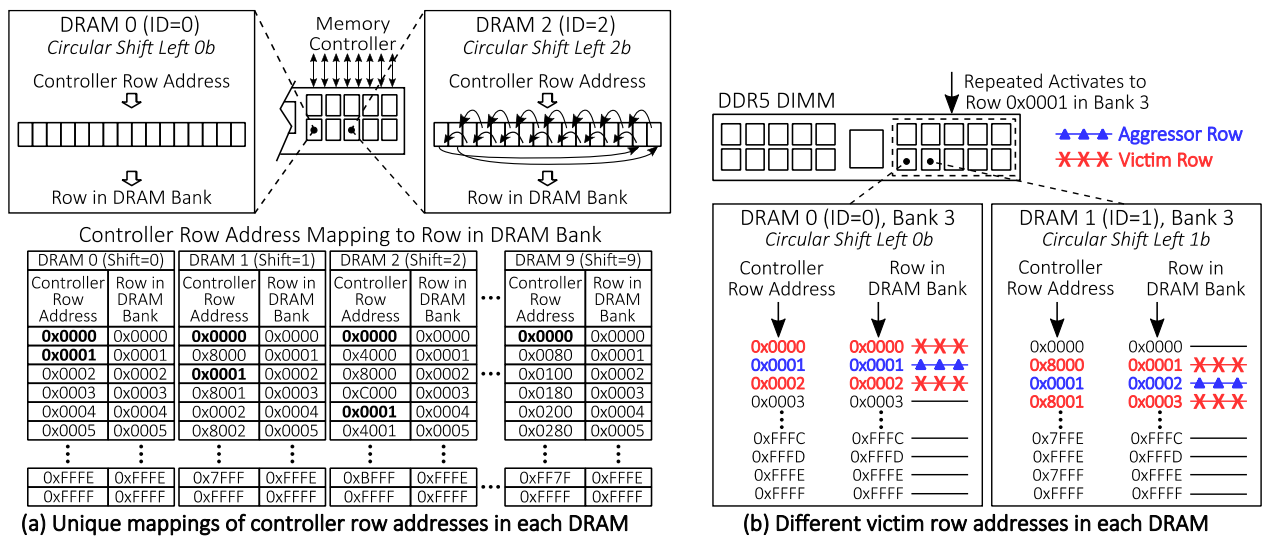

Figure 5: Example RAMPART row address mappings producing unique neighbors, so RowHammer attacks have different victim addresses in each DRAM. (a) Circular left shifts of controller row addresses based on unique DRAM IDs are shown. The tables at the bottom show how controller row addresses map to internal bank rows in each DRAM. Row addresses 0x0000 and 0x0001 are bolded to show increasing separation with larger shifts. (b) Hammering controller row address 0x0001 flips bits in controller row addresses 0x0000 and 0x0002 in DRAM 0, but controller row addresses 0x8000 and 0x8001 in DRAM 1. A subsequent read to controller row address 0x0000 sees errors only from DRAM 0 that can be corrected with SDDC ECC.

The benefit of this mapping is shown in Figure 5b. If controller row address 0x0001 is successfully hammered, different controller row addresses will be victims in each of the other DRAMs. A subsequent read to controller row address 0x0000 will only see errors from DRAM 0, as controller row address 0x0001 is not neighbors with controller row address 0x0000 in any other DRAM. This mapping confines bit flips to a single DRAM for any victim row address, allowing ECC to repair these bit flips.

Figure 5a shows a mapping in which no two controller row addresses are *immediate* neighbors in more than one DRAM. This approach also works when victims can be 2 or more rows away from the aggressor (blast radius ≥ 2), for example using larger shifts (e.g., 2-bits×ID) if the address space is large enough, and/or with other mapping functions. One challenge is that repaired rows are mapped uniquely to spare rows, making it difficult to guarantee row addresses have unique neighbors in each DRAM. In many cases, DRAMs have substantial numbers of unused spare rows, allowing repaired rows to be shielded from other rows by placing unused spare rows between them. DRAMs can be binned to identify those with sufficient unused spare rows so that remapping can guarantee all rows, even repaired ones, have unique neighbors and/or unused rows for shielding. If all repaired rows cannot be spaced sufficiently far apart, the vast majority of rows (including many repaired rows) will have unique neighbors, confining errors in these rows to one DRAM. The locations of vulnerable rows will vary across DIMMs, and attacks targeting these repaired rows may work on one DIMM but will likely fail on others. Because DRAMs already implement row address mapping on repaired rows, RAMPART's remapping function is an incremental design change that can be implemented with no additional latency.

## 4. Correcting Bit Flips with SDDC ECC

As data rates rise and process geometries shrink, memory systems are more prone to errors. In DDR5, integrity and reliability are improved using a layered ECC approach combining protection in the DRAM [6] with protection at the system-level. DDR5 DRAMs can detect single-bit errors in 128b blocks of data, returning corrected data on a read operation. An important concern is *error accumulation*, in which multiple single-bit errors occur in the same 128 bits of data over time. This results in undetectable errors and *silent data corruption* (SDC). Error accumulation can be prevented with *Error Check and Scrub* (ECS) operations that correct single-bit errors during all bank refresh and self-refresh.

Multi-bit errors can be corrected at the system level (in the controller) using additional DRAMs per rank. At the system-level, multiple bits are aggregated into *symbols*, some holding data and some holding *check symbols* for data integrity. *Reed-Solomon* (RS) coding can correct one or more symbol errors across the data and check symbols, allowing any number of bit errors within an erroneous symbol to be corrected.

DDR5 server DIMMs can provide system-level ECC by using 8 x4 DRAMs per rank for data symbols, and up to 2 additional x4 DRAMs per rank for check symbols. DIMMs that use 2 additional DRAMs per rank can implement *Single Device Data Correction* (SDDC, also called *chipkill*), a method commonly used today [57] that can correct any number of bit errors from a single DRAM. SDDC allows a system to keep running even if a DRAM fails. DIMMs can instead use 1 additional DRAM per rank, trading off a lower level of ECC protection for reduced cost. DIMMs can also forgo additional ECC DRAMs to minimize cost.

Figure 6a shows an SDDC configuration with 4b symbols using all bits from the same burst of a DRAM. RS(10,8) coding can be used on the 10 symbols to correct any number of bit errors in one symbol. Two symbols in error can be detected (but not corrected) 47% of the time, while 53% of the time they cannot be detected at all [60]. And three or more symbols in error can be detected (but not corrected) only 41% of the time, while 59% of the time they cannot be detected. Figure 6b shows an SDDC configuration with 16b symbols formed from all bits of the burst length on each data pin. RS(40,32) coding on these 40 symbols enables any number of bit errors in up to 4 symbols to be corrected. The system is protected when an entire DRAM fails, or when up to 4 symbols across the DRAMs are in error. ECC with larger symbols is more robust, but slower and more complex. With RS(40,32) and 16b symbols, the SDC probability with 5 to 8 symbol errors is below $5.0 \times 10^{-15}$, and with 8b symbols is below $2.2 \times 10^{-5}$.

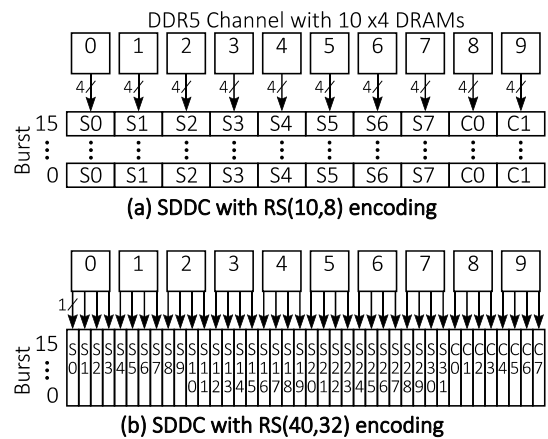

Figure 6: The mapping of bits to symbols in (a) supports SDDC ECC, allowing the system to keep running if one DRAM fails or all bits from one DRAM are in error. The mapping in (b) uses more symbols to provide SDDC ECC and the ability to correct up to 4 symbol errors.

When the controller corrects an error, a *demand scrub* operation is initiated that writes the corrected data back to the DRAMs, clearing any single-bit and multi-bit errors stored in them. *Patrol scrub* [42] is used today (typically once to a few times per day depending on the desired level of memory system reliability) to proactively read all DRAM locations and write data back to memory, correcting any errors. Patrol scrub runs in the background, with memory accesses spread out over the patrol scrub interval, and the overhead is small in modern servers. Running patrol scrub once per day on a 1TiB DDR5-5600 memory channel transfers 2TiB of read and write data, consuming 25.5MB/s or 0.057% of the peak memory bandwidth and memory power.

Our solution combines RAMPART with SDDC and patrol scrub to correct bit flips caused by one successful RowHammer attack, as errors at any victim row address are confined to a single DRAM. A variety of SDDC methods can be used, including those in Figure 6. The only ECC requirement is that the complete corruption of data from a single DRAM can be corrected. Using the example in Figure 6b, if an attack has been successful each victim row address will have at most four symbols in error, one for each of the 4 pins of the DRAM with the victim row address. If any of these victim rows is accessed, SDDC will detect and correct bit errors and write corrected data back to the DRAM, healing the data and reversing the effects of the attack. Irreparable damage to data requires a second successful attack on the same victim row address, but in a different DRAM. If the victim row address is accessed (e.g., read by an application or via patrol scrub) before the second attack succeeds, the data will be corrected and written back to memory, thwarting the attack. Referring to Figure 5b, a RowHammer attack attempting to flip bits in controller row address 0x0000 needs to successfully hammer row address 0x0001 to flip bits in DRAM 0, and row address 0x8000 to flip bits in DRAM 1. If data at victim row address 0x0000 is read before the second attack completes, bit flips from the first attack will be corrected with SDDC. RAMPART and SDDC can also be used together with other RowHammer defense mechanisms to provide an added layer of resilience.

DIMMs with 1 additional DRAM per rank instead of 2 provide a lower level of ECC protection at a reduced cost. A configuration similar to that shown in Figure 6b, but with 9 DRAMs and a similar mapping of bits to symbols, can implement 2-symbol correction using RS(36,32) coding. Up to 2 erroneous symbols can be corrected, providing protection against RowHammer attacks that only flip bits in up to 2 symbols in the same DRAM. With RS(36,32) and 16b symbols the SDC probability with 3 to 8 symbol errors is less than $1.5 \times 10^{-7}$, and with 8b symbols it is less than $9.5 \times 10^{-3}$.

Previous studies [7] have shown that RowHammer attacks often do not flip all bits in a row, instead affecting a smaller number of bits. Using 9 DRAMs per rank with RS(36,32) encoding can provide partial protection against attacks for DRAMs with a small number of susceptible bits per row, but cannot match the stronger reliability that 10 DRAMs per rank can achieve with RAMPART and SDDC. There are no limits on the number of bit flips that can be corrected with RAMPART and SDDC as a result of a single successful RowHammer attack. With 10 DRAMs per rank, *one successful attack can flip every bit in the neighboring rows and RAMPART coupled with SDDC can correct these errors.* To cause uncorrectable errors, a second successful attack is needed before errors from the first attack are detected and corrected.

## 5. Minimizing the Impact of RowHammer Mitigation Mechanisms

A key objective for RowHammer mitigation and prevention mechanisms is to increase the memory system's resistance to bit flips while minimizing any impact to performance. RAMPART and SDDC increase resistance to data corruption in server memory systems by requiring two successful attacks to cause uncorrectable memory errors. In this section we discuss BRC-VL, a variation of DDR5's BRC mitigation feature that improves performance by reducing mitigation overhead.

RAMPART can be used together with DDR5 BRC as well as with many different RowHammer tracking and sampling techniques. Numerous methods [2][21][27][30][38] have been proposed, with these approaches varying in area, complexity, and performance impact. Tracking methods identify aggressor rows based on historical accesses. Some methods like Misra-Gries [30] provide a guarantee that aggressor rows will be identified if enough candidate rows are tracked. As HC continues to fall and the number of banks and channels continues to rise, SRAM and Content Addressable Memory (CAM) storage requirements grow and can make tracking techniques challenging or impractical to implement [32][37][44]. Another concern is that future attacks may be developed that evade current algorithms.

An alternative to algorithmic tracking is random sampling, which selects aggressor rows probabilistically. Random sampling can be effective at minimizing the probability of success for RowHammer attacks and can avoid risks associated with algorithmic tracking. These methods are also simpler to implement, but do not provide certainty that an aggressor row will be identified. Instead, random sampling methods mitigate RowHammer attacks by making it statistically unlikely that an attack will succeed. PARA [8] is a well-known example of a random sampling approach.

DDR5 DRAMs support *Directed Refresh Management* (DRFM) and *Bounded Refresh Configuration* (BRC) [6], enabling controller-driven probabilistic sampling methods like PARA to be implemented (PARFM [38] uses DRFM to implement PARA). DRFM allows the controller to select a *target row address* (e.g., row R) by setting a dedicated bit in commands that precharge a row, and BRC sets the number of rows that are refreshed. The DRAM stores the target row address and refreshes its victim rows (rows R±1) when an RFM command is received. Like Refresh, RFM commands can affect all banks (RFMab), or the same bank number in each bank group (RFMsb). BRC combats victim-focused mitigation attacks by refreshing neighbors of victim rows (rows R±2), but at a lower frequency called the *refresh ratio* (or *ratio*) known only by the DRAM. After some number of activates to a bank (the *Rolling Accumulated ACT Initial Management Threshold*, RAAIMT), an RFM command is issued by the controller. The DRAM will *always* refresh one set of rows (e.g., rows R±1), and *sometimes* refresh a neighboring set of rows (e.g., R±2). The controller must allocate enough time to refresh all four rows, even though the DRAM may only refresh two rows. DRAM bandwidth is wasted, and subsequent transactions can incur additional latency, hurting system performance. Multiple BRC *support levels* are defined (our example describes BRC2), enabling different numbers of neighboring rows to always be refreshed based on RowHammer susceptibility. Although our discussion assumes hammering row R only affects immediately neighboring victim rows R±1, we note that this approach extends to larger numbers of victim rows.

We propose *Bounded Refresh Configuration with Victim Levels* (BRC-VL), a BRC variation that is compatible with DDR5 DRFM, is simple to implement, and improves system performance by reducing RFM overhead. When coupled with RAMPART, SDDC, and probabilistic sampling, BRC-VL performs well at mitigating both traditional and victim-focused mitigation attacks. BRC-VL differs from BRC by having the controller instruct the DRAM to refresh *either* the victim rows (*level 1 victims* R±1) *or* the neighbors of the victim rows (*level 2 victims* R±2). The advantage of refreshing either the level 1 or level 2 victims is reduced RFM time and improved system performance, as only two rows are refreshed. Level 1 victims are selected with higher probability than level 2 victims, mirroring the approach used in BRC. The tradeoff is a small increase in the probability of success of a single traditional RowHammer attack, because (i) level 1 victims are not refreshed with 100% probability, and (ii) if level 2 victims are selected and refreshed, the level 1 victims will be impacted twice (a situation we refer to as *double hammering*). But as we show in Section 6, this small increase in the probability of success of a single attack is more than compensated for by the need for two successful attacks to irreparably damage data when BRC-VL is combined with RAMPART and SDDC. At the system level, the combination of RAMPART and shorter BRC-VL RFMs results in higher application performance and improved resistance to data corruption compared to DDR5 BRC. Like BRC, BRC-VL RFM commands can be directed at all banks or the same bank in each bank group. We describe BRC-VL with two victim levels, and note the approach can extend to more levels and can also refresh larger numbers of rows. More levels may be needed depending on system parameters and the susceptibility of the DRAM to RowHammer.

BRC-VL selects the *victim level* (level 1 or level 2 victims) based on probabilities that work well against a range of attacks. Since RFMs are issued after RAAIMT activates to a bank, a high-frequency attack that activates only one row (row R) will hammer it RAAIMT times before the controller selects it. If the RFM refreshes level 1 victims (rows R±1) once after every RAAIMT activates to the aggressor row, these refreshes will hammer the level 2 victims (rows R±2) but at a lower frequency, 1/RAAIMT, compared to the aggressor row. When an RFM is issued, we refresh level 1 victims with a probability of (RAAIMT-1)/RAAIMT, and the level 2 victims with a probability of 1/RAAIMT.

A sample implementation of the BRC-VL selection logic was designed for a 32-bank system with 2 victim levels and consists of 4 parts: (i) a random number generator, (ii) target row selection, (iii) Bank Activate Counters (BACs) in the controller that track the number of activates to each bank, and (iv) signaling the need for an RFM command. Many options exist for random number generation, and a full discussion is beyond the scope of this paper. Our design uses a 16-bit LFSR that generates a single random bit every clock cycle. The 16-bit random number is sampled at the beginning of each *RAAIMT window* (next RAAIMT activates to a bank) and is used to select the target row. The number of clock cycles per window can vary due to scheduling, making it difficult to predict the sampled number. The maximum supported RAAIMT value is 256, and up to 8 bits from the 16-bit random number are used to select the target row from the RAAIMT window. Up to 8 of the remaining bits are used to select the victim level. Level 2 victims are selected if the bits match a specific value, otherwise level 1 victims are chosen. When the target row is precharged, the DRAM saves the row address and victim level for use with the next RFM command. For large memory systems with hundreds of banks or more, area efficiency can be improved if multiple banks share a random number generator. Our design shares one random number generator across all 32 banks, with each bank receiving a unique arrangement of bits from the LFSR. A tracking block monitors the BACs and indicates when an RFM operation is needed.

Our 32-bank design was synthesized in TSMC's 7nm process. Based on raw synthesis results, and assuming 70% area utilization and conservative routing, our design reaches a speed of 2.85GHz in an area of 3910um$^2$, or roughly 51K NAND2 gates. For a server with 1024 banks, the total area required is only 0.1251mm$^2$. The small size illustrates a key advantage of probabilistic selection methods like PARFM that are coupled with BRC or BRC-VL.

## 6. Evaluation and Analysis

In this section, we analyze the effectiveness of RAMPART and BRC-VL when coupled with controller-driven probabilistic sampling. We compare the theoretical resistance and performance impact of PARFM with BRC (a DDR5 implementation of PARA) against PARFM with BRC-VL, RAMPART, and SDDC. When BRC-VL is combined with RAMPART and SDDC, systems benefit by (i) having stronger resistance to data corruption from RowHammer attacks due to needing two successful attacks to cause irreparable data damage, (ii) having more available memory bandwidth resulting from shorter RFM operations, and (iii) enabling better application performance due to higher memory bandwidths and reduced RFM latencies. We explore these benefits in the following sections. For the rest of this paper, we assume that BRC-VL is combined with RAMPART and SDDC unless noted.

## 6.1 Data Corruption Probability

We first analyze BRC-VL's data corruption probability in the same manner used to evaluate PARA [8], with an attack that activates an aggressor row exactly HC times per refresh interval. We assume an RAAIMT window with N activates, and at most one aggressor row activate per window (i.e., some windows may have none in them). In windows that have an aggressor row activate in them, the probability of randomly selecting the aggressor row is (1/N), and the probability of refreshing level 1 victims is (N-1)/N. The probability of refreshing the correct victim row in an RAAIMT window is then $(N-1)/N^2$, and the number of victim row refreshes can be treated as a binomially-distributed random variable with parameters $B(HC,(N-1)/N^2)$. Bit flips occur in the level 1 victim rows only if they are never refreshed in any of the RAAIMT windows during the refresh interval. The probability that this occurs is $(1-(N-1)/N^2)^{HC}$.

The *effective hammer count* HCe in BRC-VL needs to be considered due to *double hammering* that can happen to rows R±1. If row R is activated and its level 2 victims (rows R±2) are refreshed, rows R±1 are impacted twice. The probability of this happening in any RAAIMT window is $1/N^2$. For the smallest RAAIMT window (N=16) in our study, this probability is 1/256. For our most susceptible DRAMs (HC=1000), on average about 4 double hammers occur across 1000 RAAIMT windows, reducing HC from 1000 to an effective value of HCe=996. This minor decrease in hammer count is more than compensated for because RAMPART with SDDC requires two successful attacks to cause irreparable damage to data. For two attacks to be successful, the level 1 victims of each attack need to avoid being refreshed 996 times on average, for a total of 1992 missed refreshes. Although it is possible in this example for 1000 consecutive double hammers to occur (the minimum number of aggressor activates to corrupt data when BRC-VL is coupled with RAMPART and SDDC), the probability of this happening is vanishingly small at $(1/256)^{1000}$.

Table 1 shows the probability of one successful attack in a 32ms refresh interval and over one year for two different HC values. The calculations use the corresponding HCe values shown. Also shown is the probability of data corruption due to two successful attacks in two consecutive refresh intervals (64ms), and over the course of a year. The smaller HCe values change the probabilities by less than one order of magnitude and show that double hammering impact is minimal for the smallest RAAIMT values we study.

**Table 1: BRC-VL Attack Success Probability Analysis**

| 1 successful attack | | |
|---|---|---|
| RAAIMT=16 | HC=1000 (HCe=996) | HC=3000 (HCe=2988) |
| 32ms | 7.6×10$^{-27}$ | 4.4×10$^{-79}$ |
| 1 year | 7.5×10$^{-19}$ | 4.4×10$^{-71}$ |

| 2 successful attacks (data corruption) | | |
|---|---|---|
| RAAIMT=16 | HC=1000 (HCe=996) | HC=3000 (HCe=2988) |
| 64ms | 5.8×10$^{-53}$ | 2.0×10$^{-157}$ |
| 1 year | 5.6×10$^{-37}$ | 1.9×10$^{-141}$ |

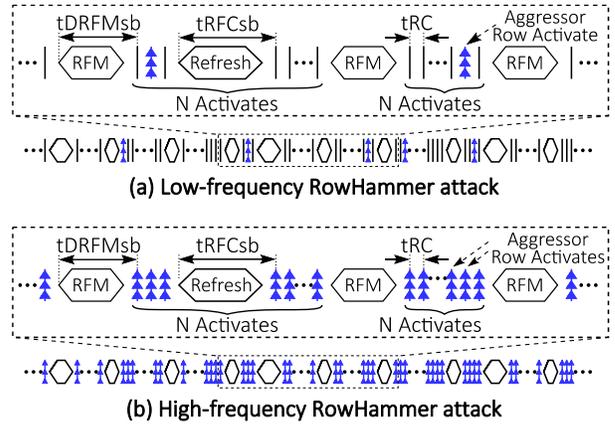

Figure 7: RowHammer Attack Models. (a) Low-frequency attacks hide aggressors in a sea of decoy activates. (b) High-frequency attacks enable victim-focused mitigation attacks.

## 6.2 RowHammer Attack Models

More aggressive attacks that hammer rows at a higher frequency than the one discussed in Section 6.1 improve the chances of flipping bits over time. We use the two RowHammer attack models shown in Figure 7 to evaluate the resistance of DDR5 BRC and BRC-VL: (i) a *low-frequency attack* that randomly mixes 1 hammer and N-1 decoy activates per RAAIMT window, issuing them back-to-back to the same bank, and (ii) a *high-frequency attack* that issues back-to-back activates to the same row in the same bank. Low-frequency attacks are difficult to identify with algorithmic tracking [2]. High-frequency attacks are easier to track but can promote victim-focused mitigation attacks. The aggressor row activates in Figure 7 can be to the same address (a *single-sided attack*) or to multiple addresses affecting the same victim row (a *multi-sided* attack). The probability of selecting an aggressor row depends on how many aggressor activates occur in an RAAIMT window, not on the number of sides in the attack. Both attacks are useful for studying bounds on the susceptibility to attacks aimed at victim rows, independent of the number of sides attacked.

We model BRC with a ratio of 1/RAAIMT to match the approach used by BRC-VL, which makes its resistance to both attacks similar mathematically. For the traditional attack, the probability of selecting aggressor row R in an RAAIMT window of size N is 1/N, and rows R±1 are always refreshed so the probability the attack is stopped is 1/N in any RAAIMT window. For a victim-focused mitigation attack the probability of selecting aggressor row R is 1, and rows R±2 are refreshed with probability 1/N, so the probability an attack is stopped in an RAAIMT window is also 1/N.

## 6.3 Data Corruption Probability vs. Time

Markov models were created to statistically analyze the resistance to data corruption over time of PARFM with BRC and PARFM with BRC-VL, RAMPART, and SDDC using these more aggressive attack models. The primary difference is that one successful RowHammer attack is needed to cause uncorrectable errors in BRC, while RAMPART and SDDC strengthen BRC-VL by requiring two successful RowHammer attacks to cause uncorrectable errors. We assume the controller is secure and its target row choices cannot be observed by an attacker. Our BRC-VL model also assumes an *oracle attacker* that knows when a first attack has been successful and switches to a different aggressor row to attack the same controller row address in a different DRAM. It is unclear how to achieve this, as RAMPART and SDDC will ensure that reading data at the victim row address will correct any bit flips caused by the first attack (requiring the attack to start over), but we study this case as it provides a worst-case time bound on the susceptibility to two successful attacks. Because BRC-VL with RAMPART and SDDC requires two attacks to corrupt data compared to just one attack with BRC, to first order the probability of data corruption at any time with this combination is the square of the probability of data corruption with BRC. Our models evaluate the effectiveness of both methods over one year of continuous attacks to a single DRAM bank, using DDR5-5600 timing parameters shown in Table 2. HC values of 1000 to 3000 are used, reflecting near-future possibilities [9][32]. Our BRC-VL models use four victim levels, as victim-focused RFMs for our lowest RAAIMT values can recursively induce attacks on other victims multiple rows away. These attacks can weaken RowHammer defenses if $APR \geq RAAIMT^{(VL+1)}$, where $APR$ is the number of activates per refresh period, and $VL$ is the maximum victim level. When considering the number of victim levels to support, systems should choose $VL \geq \log_{RAAIMT}(APR) - 1$.

Figure 8 illustrates the susceptibility of a DDR5-5600 DRAM bank to a traditional low-frequency RowHammer attack (shown in Figure 7a) when using PARFM and BRC. The graphs show the probability of a continuous attack successfully evading PARFM's random selection enough times to cause bit flips – HC times in a row across RAAIMT consecutive windows. The graphs plot the probability of at least one attack being successful over a year of continuous attacks as RAAIMT and HC vary. Susceptibility falls as HC increases and as RAAIMT is reduced. The former indicates the DRAM has a stronger inherent resistance to RowHammer, while the latter enables PARFM to issue more frequent RFM operations designed to thwart these attacks. The vulnerability of a bank can be adjusted across many orders of magnitude by changing RAAIMT, but the graphs also show that for a given range of RAAIMT values, the range over which vulnerability can be adjusted narrows as HC falls.

Table 2: DDR5-5600 Model Parameters

| Param | Value | Description |
|---|---|---|
| tRC | 46.4ns | Row cycle time |
| tREF | 32ms | Refresh period |
| tREFIsb | 487.5ns | Same bank refresh interval |
| tRFCsb | 130ns | Same bank refresh cycle time |
| tDRFMsb | 240ns (BRC) 130ns (BRC-VL) | Same bank DRFM duration |
| HC | 1000-3000 | Hammer Count |
| RAAIMT | 16-100 | RAAIMT window size |

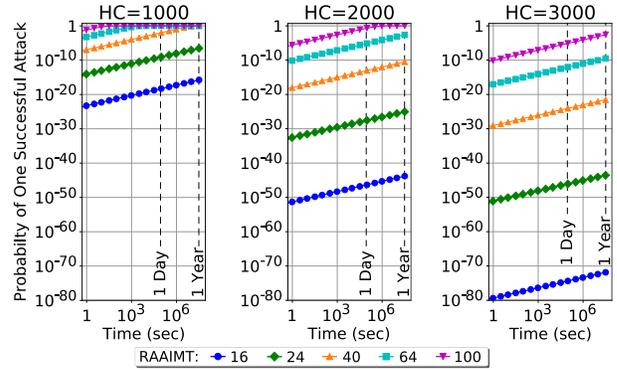

Figure 8: Resistance of a DDR5-5600 DRAM bank to a traditional low-frequency RowHammer attack when using PARFM with BRC. Systems with higher HC values and lower RAAIMT values are more resistant to RowHammer attacks. Resistance can be adjusted across many orders of magnitude by changing RAAIMT, but the range over which vulnerability can be adjusted narrows as HC falls.

Figure 9 shows that for the most vulnerable DRAMs simulated (HC=1000, 1500), issuing RFMs infrequently (RAAIMT=100) leads to near-certain data corruption with traditional attacks in just minutes when using BRC. The same is true for victim-focused mitigation attacks, which are not shown. And even though RAMPART with SDDC improves the resistance of BRC-VL, data corruption is still likely to occur in less than an hour. For the most vulnerable DRAMs, patrol scrub does not run often enough in modern servers to allow

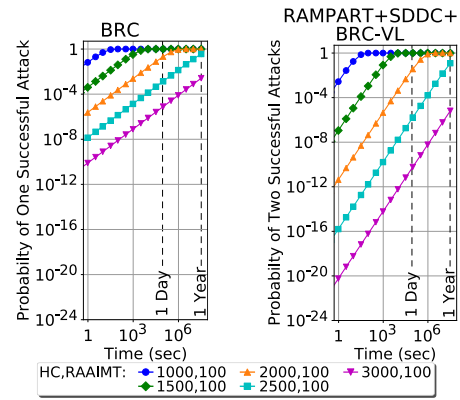

Figure 9: For the most vulnerable DRAMs studied, infrequent RFMs relative to HC lead to near-certain data corruption under traditional attacks. RAMPART and SDDC strengthen resistance to data corruption, but for HC=1000 and RAAIMT=100, irreparable data damage can still occur in minutes.

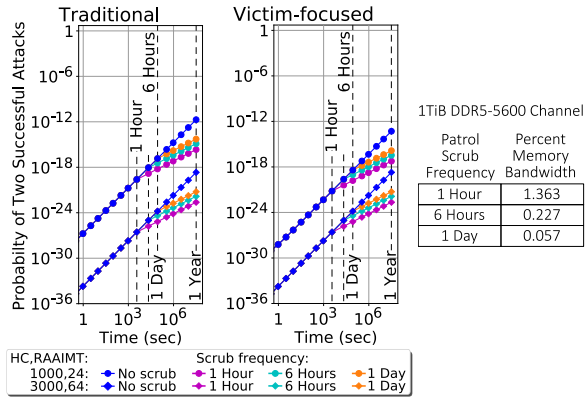

Figure 10: Impact of combining RAMPART, SDDC and BRC-VL with patrol scrub run every hour, every 6 hours, and every day. Patrol scrub reverses the damage from a single RowHammer attack when used with RAMPART and SDDC, increasing resistance to attacks. Running patrol scrub more often improves resistance, but consumes more memory bandwidth.

RAMPART with SDDC to repair damage from the first attack before a second attack is likely to succeed.

Figure 10 illustrates the impact of running patrol scrub together with RAMPART, SDDC, and BRC-VL. When used with RAMPART and SDDC, patrol scrub can reverse the effects of a single successful RowHammer attack, increasing memory system resistance. Once patrol scrub begins, the probability of data corruption (due to two successful attacks) follows a shallower curve, improving resistance by several orders of magnitude over a year of continuous attacks compared to not using patrol scrub. More frequent patrol scrubs provide higher resistance to attacks, but consume more memory bandwidth, reflecting the tradeoff between system resistance and performance impact.

Issuing RFMs more frequently dramatically improves resistance to both types of attacks, as shown in Figure 11a and Figure 11b. Both figures compare the effectiveness of three mitigation methods: (i) PARFM with BRC, (ii) PARFM with BRC-VL, and (iii) PARFM with RAMPART, SDDC, BRC-VL and patrol scrub run once per day. PARFM with BRC-VL refreshes slightly fewer rows on average compared to PARFM with BRC, trading off a small increase in data corruption probability against lower mitigation overhead. The difference in susceptibility for these two methods is within an order of magnitude over the simulated duration of attacks. These figures also show that RAMPART with SDDC more than compensates for BRC-VL's small increase in RowHammer susceptibility by requiring two successful attacks to corrupt data before errors from the first attack are corrected. This essentially squares the probability of data corruption compared to using BRC-VL alone. The use of patrol scrub further improves resistance to attacks compared to BRC.

Table 3 summarizes the resistance of BRC and BRC-VL with RAMPART, SDDC, and patrol scrub after one day and one year when HC=1000 and RAAIMT=24. The probability of data corruption for BRC-VL with RAMPART and SDDC is roughly the square of the probability when using DDR5 BRC before patrol scrub begins, again due to requiring two successful attacks. After patrol scrub starts, there are further gains in resistance compared to BRC. For the range of parameters and attacks in Figure 11, BRC-VL with RAMPART, SDDC, and

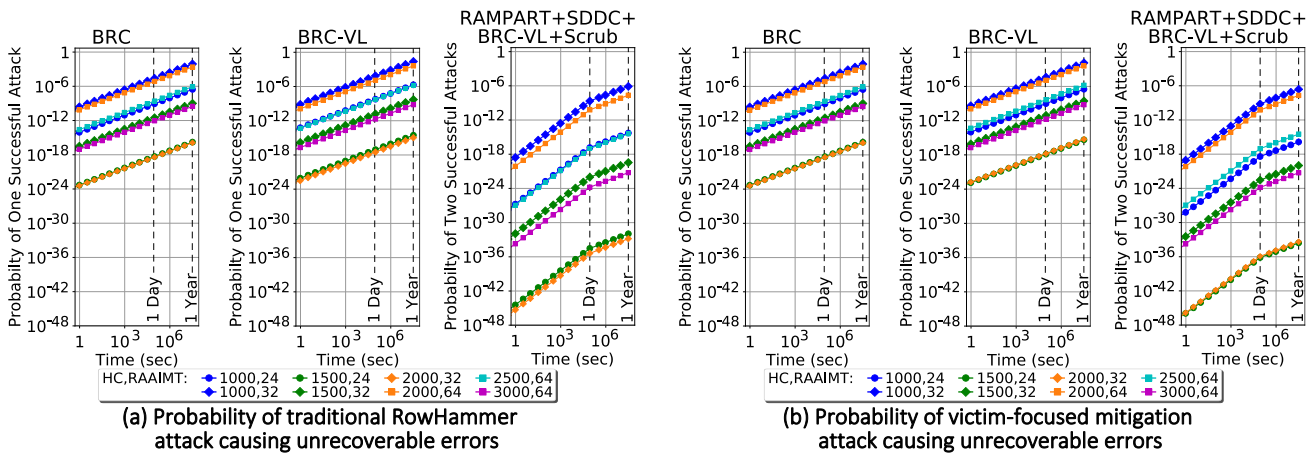

(a) Probability of traditional RowHammer attack causing unrecoverable errors

(b) Probability of victim-focused mitigation attack causing unrecoverable errors

Figure 11: Probability of data corruption over time in one DDR5-5600 DRAM bank under continuous attack. The "BRC" graphs depict PARFM (a DDR5 implementation of PARA) using DDR5 BRC, while the "BRC-VL" graphs depict PARFM coupled with BRC-VL. The "RAMPART+SDDC+BRC-VL+Scrub" graphs depict PARFM coupled with RAMPART, SDDC, BRC-VL, and patrol scrub run once per day. RAMPART with SDDC, BRC-VL, and patrol scrub provides stronger resistance to data corruption than BRC and BRC-VL under (a) traditional attacks and (b) victim-focused mitigation attacks as two successful attacks are needed to irreparably damage data. The probabilities for BRC are mathematically similar in both attacks. For the most vulnerable DRAMs analyzed (HC=1000), the probability of data corruption using RAMPART with SDDC and BRC-VL is below $1.45 \times 10^{-17}$ for both attacks after one day with RAAIMT=24, an improvement of nearly 8 orders of magnitude over BRC. Across the configurations shown, RAMPART with SDDC, BRC-VL, and patrol scrub achieves between 4 and 17 orders of magnitude lower probability of data corruption after one year of continuous attacks compared to DDR5 BRC. Resistance can be further improved in all cases by reducing RAAIMT, but this also increases RFM overhead and can impact system performance.

Table 3: Markov Model Data Corruption Probability

| HC=1000 RAAIMT=24 | BRC | | RAMPART+SDDC+ BRC-VL+Scrub | |
|---|---|---|---|---|
| | 1 Day | 1 Year | 1 Day | 1 Year |
| Single Bank Attack, DDR5-5600 | | | | |
| Traditional | 7.8×10⁻¹⁰ | 2.9×10⁻⁷ | 1.4×10⁻¹⁷ | 5.3×10⁻¹⁵ |
| Victim Focused | 7.8×10⁻¹⁰ | 2.9×10⁻⁷ | 3.8×10⁻¹⁹ | 1.4×10⁻¹⁶ |
| System Attack (16 banks attacked per channel, 32 channels) | | | | |
| Traditional | 4.0×10⁻⁷ | 1.5×10⁻⁴ | 7.4×10⁻¹⁵ | 2.7×10⁻¹² |
| Victim Focused | 4.0×10⁻⁷ | 1.5×10⁻⁴ | 2.0×10⁻¹⁶ | 7.2×10⁻¹⁴ |

Table 4: Analytical Bound on Data Corruption Probability

| HC=1000 RAAIMT=24 | BRC | | RAMPART+SDDC+ BRC-VL+Scrub | |
|---|---|---|---|---|
| | 1 Day | 1 Year | 1 Day | 1 Year |
| Single Bank Attack, DDR5-5600 | | | | |
| Traditional | 2.1×10⁻⁹ | 7.5×10⁻⁷ | 1.5×10⁻¹⁶ | 5.4×10⁻¹⁴ |
| Victim Focused | 2.1×10⁻⁹ | 7.5×10⁻⁷ | 1.7×10⁻¹⁶ | 6.3×10⁻¹⁴ |
| System Attack (16 banks attacked per channel, 32 channels) | | | | |
| Traditional | 1.1×10⁻⁶ | 3.9×10⁻⁴ | 7.6×10⁻¹⁴ | 2.8×10⁻¹¹ |
| Victim Focused | 1.1×10⁻⁶ | 3.9×10⁻⁴ | 8.8×10⁻¹⁴ | 3.2×10⁻¹¹ |

patrol scrub provides between 4 and 17 orders of magnitude lower probability of data corruption after one year of continuous attacks. Periodic patrol scrubs coupled with RAMPART and SDDC provide additional protection as they can correct errors caused by the first attack, requiring the attacker to start over from the beginning. With DDR5-5600 DRAMs, 16 banks per channel can be simultaneously attacked at the rates shown in Figure 7. In addition, modern servers can have 16 or more memory channels, each of which can be attacked in parallel. These attacks are independent, and the probability of corrupted data in at least one bank across a 32-channel memory system, where 16 banks are attacked simultaneously per channel, is shown in Table 3.

We further analyzed BRC and BRC-VL coupled with RAMPART, SDDC, and patrol scrub by creating analytical models that are upper-bounds on our Markov models. For any attack, if the probability of refreshing the victim rows after an RAAIMT window is $p$, the probability of an attack succeeding after $s$ windows is $sp(1-p)^{HC} < spe^{-pHC}$ for a hammer count of $HC$. With BRC-VL, the upper bound formula is modified to account for double-hammering in traditional attacks, and the probability of an attack succeeding is bounded by $spe^{-\left(\frac{N-1}{N^2+1}\right)HC}$. If the system is subject to $k$ simultaneous attacks, both formulas are multiplied by $k$, up to a maximum of $se^{-\left(\frac{N-1}{N^2+1}\right)HC}$ due to bandwidth limitations. If multiple simultaneous attacks are nearly independent, the probability of two attacks succeeding is no more than approximately $\left(se^{-\left(\frac{N-1}{N^2+1}\right)HC}\right)^2$ per channel. Table 4 shows the resulting probabilities, which are upper bounds on the Table 3 values.

Multiple simultaneous attacks against a bank were also considered. We analyzed the resistance of PARFM with BRC against PARFM with RAMPART, SDDC, BRC-VL, and patrol scrub run once per day under one attack, and under 10 simultaneous attacks. For our RAMPART simulations, we assume the single attack scenario uses an oracle attacker, and the 10 simultaneous attack scenario uses 10 non-oracle attackers that target the same victim row address, but in different DRAMs. Each DRAM is the subject of one attack that hammers a different row address, but that has the same victim row address across all DRAMs. The goal of this attack is to have at least 2 attacks succeed, causing irreparable damage to data that SDDC cannot repair. We assume the success of separate attacks is independent, but our analysis suggests a slight anticorrelation that further helps RAMPART.

The results of this analysis are shown in Table 5. Ten simultaneous attacks against a bank increase the probability of data corruption for both BRC and RAMPART coupled with SDDC, BRC-VL, and patrol scrub. But this is not enough to overcome the requirement of two successful attacks to cause irreparable data damage when using RAMPART and SDDC.

### 6.4 Memory Subsystem Performance Impact

We created a DRAMSys4 [49]-[51] memory subsystem model to study how BRC and BRC-VL impact memory bandwidth as RAAIMT varies. Activates are tracked with Bank Activate Counters (BACs, see Section 5), and one same bank RFM (RFMsb) is issued after a BAC value reaches a value of RAAIMT for any bank in the bank group. BAC values are decremented by RAAIMT (to a minimum of 0) for banks affected by the RFMsb operation, and refreshes do not decrement the BACs. The target row and BRC-VL victim level are communicated in commands that precharge rows, as is done in DDR5 DRFM.

We model a single-channel DDR5-5600 memory system with 2 ranks and 32 banks/rank, using the timing parameters in Table 2. The controller implements a closed adaptive page policy and a 32-deep scheduling buffer. Two workloads were run: (i) a 64-byte aligned random address stream (*rand*), and (ii) a simulated RowHammer attack (*hamR*) with 20% of the transactions targeting one victim row, and the rest randomly distributed. Both use a 2:1 read-write ratio and deliver transactions at a rate that keeps the scheduling buffer full.

Figure 12 shows the bandwidth efficiency for BRC and BRC-VL as a function of RAAIMT for both workloads, comparing them to baselines with RFMs disabled. The hamR workloads have lower efficiency due to the concentration of accesses to one bank, which leads to higher bank contention. For RAAIMT>60, relatively infrequent RFM operations reduce bandwidth efficiency by less than 2% compared to the baseline for both BRC and BRC-VL, with BRC-VL having up to 1.2% better efficiency (hamR, RAAIMT=100). For RAAIMT≤60, BRC-VL's shorter RFM operations result in larger performance advantages. At RAAIMT=16, the difference in bandwidth efficiency between BRC-VL and BRC is 5.0% for rand, and 2.5% for hamR. As hammer counts fall, RAAIMT values will need to fall as well, and the results in Figure 12 illustrate that the shorter BRC-VL RFMs can improve available memory system bandwidth.

Table 5: Simultaneous Attack Data Corruption Probability

| HC=1000 RAAIMT=24 | BRC | | RAMPART+SDDC+ BRC-VL+Scrub | |
|---|---|---|---|---|
| | 1 Day | 1 Year | 1 Day | 1 Year |
| Single Bank Attack, DDR5-5600 | | | | |
| Traditional | 7.8×10⁻¹⁰ | 2.9×10⁻⁷ | 1.4×10⁻¹⁷ | 5.3×10⁻¹⁵ |
| Victim Focused | 7.8×10⁻¹⁰ | 2.9×10⁻⁷ | 3.8×10⁻¹⁹ | 1.4×10⁻¹⁶ |
| 10 Simultaneous Attacks Against 1 Bank | | | | |
| Traditional | 7.8×10⁻⁹ | 2.9×10⁻⁶ | 1.3×10⁻¹⁵ | 4.7×10⁻¹³ |
| Victim Focused | 7.8×10⁻⁹ | 2.9×10⁻⁶ | 3.4×10⁻¹⁷ | 1.3×10⁻¹⁴ |

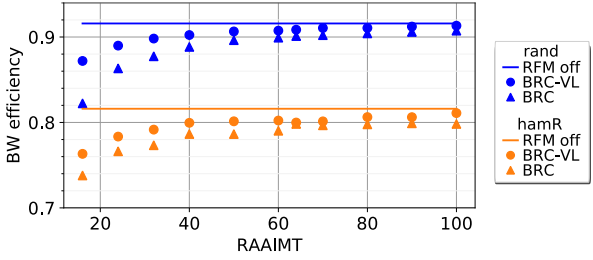

Figure 12: BRC and BRC-VL bandwidth efficiency for high-bandwidth random (rand) and simulated attack (hamR) workloads over a range of RAAIMT values. The bandwidth efficiency with RFMs disabled is also shown for comparison.

### 6.5 System-Level Performance Impact

We created a system-level model in gem5 [52][53] to analyze the performance impact of BRC and BRC-VL on various application workloads. We modified gem5 to support same bank refreshes and RFMs to match our DRAMSys4 model. The controller suspends activates to a bank when BAC values reach a limit of 2×RAAIMT. Our gem5 and DRAMSys4 models were correlated using the same configurations and workloads described in Section 6.4. Our models correlate well, with bandwidth efficiencies being within 4.7% and 1.5% on rand and hamR, respectively. The largest differences occurred at RAAIMT=16, when more RFMs were issued.

Our gem5 model uses the DRAM timings in Table 2, and the system parameters shown in Table 6. We simulate a single-core x86 CPU with a two-channel 64GiB DDR5-5600 memory system running Ubuntu 18.04.2 (kernel v4.9.186), with per-core cache sizes on par with current SoCs [59]. The SPEC2017 [54] and GAP [55] benchmarks were run, with statistics captured for the first 100 seconds of the region of interest, or for the full region of interest if runtime is shorter. Our GAP runs use a synthetic graph with degree 15 comprised of 4.2M nodes and 64M undirected edges.

Table 6: System Simulation Parameters

| Parameter | Value |
| --- | --- |
| CPU | 1 x86 core running at 3GHz |
| L1/L2 Cache Sizes | 32KiB/2MiB |
| DRAM Channels | Two 40-bit DDR5-5600 channels |
| DIMM Configuration | 2 ranks/channel, 32 banks/rank |
| Page Policy | Closed Adaptive |

Figure 13a shows the CPI impact of BRC and BRC-VL for several RAAIMT values normalized to a baseline with RFMs deactivated. A cross section of SPEC workloads with a range of working set sizes and DRAM bandwidths based on prior characterization [58] are shown. Many benchmarks in SPEC have low memory activity at these cache sizes, resulting in a geometric mean across the full SPEC suite of <1% CPI impact for BRC-VL and <1.5% for BRC in the worst case when RAAIMT=16. Results for BRC show a maximum CPI increase of 6.8% in GAP bfs and 5% in SPEC lbm_s with RAAIMT=16. Results for BRC-VL show a lower CPI impact of 2.9% in bfs (a relative difference of 3.9% compared to BRC) and 2.7% in lbm_s at the same RAAIMT value. BRC-VL's shorter RFMs reduce mitigation impact, especially at the lowest RAAIMT settings and in more memory intensive workloads. The overhead of RFMs is harder to hide in these cases, which occur up to 15K times (lbm_s) per 32ms refresh period. The performance gap between BRC-VL and BRC decreases with larger RAAIMT values and has negligible impact when RAAIMT=100.

Figure 13b illustrates how the change in bandwidth efficiency and total bandwidth vary with RAAIMT for the two benchmarks with the highest CPI impact from RFM operations. Results are normalized to the case where RFMs are disabled. The graphs show similar trends as the high bandwidth workloads of Figure 12, but bandwidth efficiency impact is lower as these programs are less memory intensive. For BRC, efficiency drops by up to 0.39% and 0.27% on lbm_s and bfs, respectively. The percentage of DRAM bandwidth achieved decreases more significantly however, with BRC seeing reductions of up to 6.7% and 5% for bfs and lbm_s. CPI increases are directly related to this reduction in bandwidth.

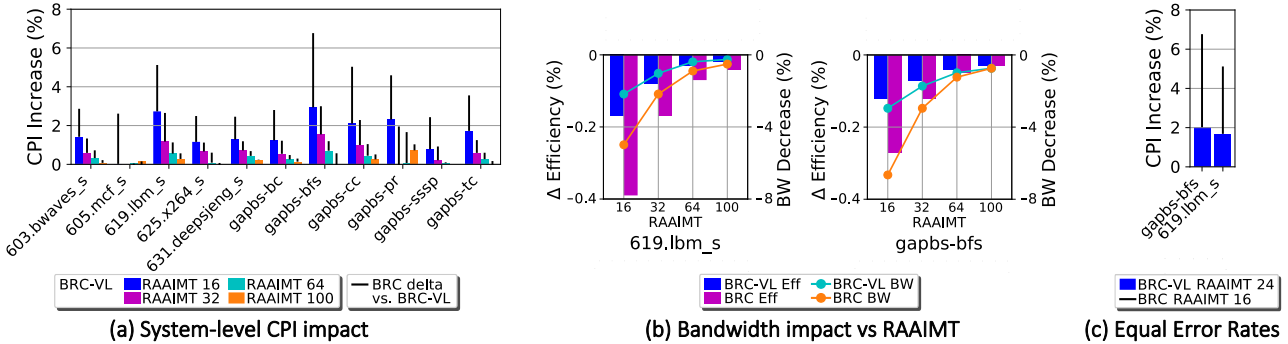

(a) System-level CPI impact    (b) Bandwidth impact vs RAAIMT    (c) Equal Error Rates

Figure 13: (a) Impact of BRC and BRC-VL on CPI, normalized to the case with RFMs disabled. (b) Bandwidth impact of RowHammer mitigation on two of the most sensitive applications in our study. (c) Improved CPI benefit of BRC-VL when systems target a fixed error rate of less than $1.0 \times 10^{-14}$ after one year of continuous attacks against a DRAM bank.

BRC-VL also benefits systems targeting a particular resistance to RowHammer data corruption. If a system with HC=1000 targets a probability of data corruption in a bank of under $1.0 \times 10^{-14}$ after one year when subject to a traditional low-frequency attack, BRC can achieve this with RAAIMT=16 (probability of $1.7 \times 10^{-16}$). BRC-VL can meet this target with RAAIMT=24 (probability of $5.3 \times 10^{-15}$). Figure 13c shows the relative CPI difference of BRC-VL (RAAIMT=24) and BRC (RAAIMT=16) for the most sensitive applications in Figure 13a. The results show that BRC-VL reduces CPI impact by 4.8% on bfs, and by 3.4% on lbm_s.

## 7. Related Work

Breaking the spatial relationship between aggressor rows and victim rows has been proposed in several controller-based and DRAM-based mitigation and prevention studies. Controller-based approaches track frequently accessed rows and remap them to break the spatial relationship between aggressors and victims [32][45], using area for the tracking structures. Random sampling can eliminate tracking structure storage, but without per-DRAM remapping a successful attack can flip bits in a victim row address in multiple DRAMs, overwhelming ECC correction capabilities. RAMPART and SDDC can be used with DDR5 RFM-compatible probabilistic sampling to save area and correct bit flips from a single successful attack. RAMPART can also be used as an added protection layer with other tracking methods.

DRAM-based mitigation methods provide an alternative to controller-based methods. The remapping method in [42] analyzes a set of DRAMs to identify bit cells susceptible to RowHammer, then applies statistical analysis to remap rows differently in each DRAM to minimize (but not prevent) uncorrectable errors due to a successful attack. Each DRAM in a rank has a unique design with different decoder contacts, making this a more expensive solution. DRAMs may also need periodic re-examination to address changes in bit cell reliability due to process drift. SHADOW [44] dynamically remaps rows in the DRAM during RFM ops, using one spare row per subarray. The latency of accesses that require activates increases as remapping information is retrieved before the activate is issued. Panopticon [37] adds per row activation counters by modifying two MATs per subarray. Its TRR-based mitigation is less efficient than remapping schemes in DRAMs with a larger blast radius [44]. RAMPART differs from these DRAM-based methods by enabling bit flips from a single attack to be repaired when coupled with SDDC, using programmable mappings in a single DRAM design with no mat changes, having no latency impact for remapping, and accommodating repaired rows.

A DIMM-based remapping technique is described in [43] that remaps addresses with logic in the RCD and 6 extra CA pins. Different DRAMs in a rank connect to a subset of these additional pins to create address permutations. One concern is that internal DRAM mappings are not public and can conflict with DIMM-based mappings, negating the protection for certain rows and for repaired rows. RCDs also have multiple CA bus interfaces [56], with each requiring additional pins and routing space that can increase cost. RAMPART does not require changes to the RCD or DIMM and can accommodate repaired rows.

## 8. Conclusion

RowHammer continues to be a challenging problem, one that process technology and circuit improvements alone are unlikely to solve in the future. In this paper, we introduce RAMPART, a novel DRAM row address remapping method that confines bit flips from a successful RowHammer attack to a single DRAM for each victim row address. When RAMPART is coupled with SDDC ECC, a method commonly used today, a single successful attack can flip every bit in the victim rows and the system can correct them and continue running. A second successful attack is required to irreparably damage data before errors from the first attack are detected and corrected, either through normal memory accesses or with patrol scrub. We also introduce BRC-VL, a variation of DDR5 BRC with reduced overhead that enables higher system performance and that can be used together with RAMPART.

The combination of RAMPART, SDDC, and BRC-VL provides stronger resistance to data corruption from RowHammer attacks and enables higher system performance compared to DDR5's BRC mitigation. We compare DDR5-5600 memory systems using RAMPART coupled with SDDC, BRC-VL, and patrol scrub against similar memory systems using DDR5 BRC. We study their impact on memory channel efficiency as well as application performance. For the RowHammer attacks and range of parameters discussed, our models show the combination of RAMPART, SDDC, BRC-VL, and patrol scrub run once per day can reduce the probability of data corruption by 4 to 17 orders of magnitude after one year of continuous attacks. Across the benchmarks and parameters studied, BRC-VL's lower overhead results in a smaller CPI increase versus DDR5 BRC when both are compared to a system with RFMs disabled. Our study shows that the difference in relative CPI impact can reach 3.9% at the same RAAIMT value and can be higher when targeting equal resistance to data corruption.